# A Novel Jamming Attacks Detection Approach Based on Machine Learning for Wireless Communication

Youness Arjoune, Fatima Salahdine, Md. Shoriful Islam, Elias Ghribi, Naima Kaabouch

School of Electrical Engineering and Computer Science
University of North Dakota, Grand Forks, United States

*Abstract*—Jamming attacks target a wireless network creating an unwanted denial of service. 5G is vulnerable to these attacks despite its resilience prompted by the use of millimeter wave bands. Over the last decade, several types of jamming detection techniques have been proposed, including fuzzy logic, game theory, channel surfing, and time series. Most of these techniques are inefficient in detecting smart jammers. Thus, there is a great need for efficient and fast jamming detection techniques with high accuracy. In this paper, we compare the efficiency of several machine learning models in detecting jamming signals. We investigated the types of signal features that identify jamming signals, and generated a large dataset using these parameters. Using this dataset, the machine learning algorithms were trained, evaluated, and tested. These algorithms are random forest, support vector machine, and neural network. The performance of these algorithms was evaluated and compared using the probability of detection, probability of false alarm, probability of miss detection, and accuracy. The simulation results show that jamming detection based random forest algorithm can detect jammers with a high accuracy, high detection probability and low probability of false alarm.

## I. INTRODUCTION

5G is expected to substitute previous generations of cellular networks in the near future, promising higher throughput and lower latency [1] thereby enabling applications such as "self-driving" cars, Internet of Things, E-health services, augmented reality, and smart cities. As a result, billions of wireless devices are expected to be connected to the internet. Like the existing networks, 5G is vulnerable to the cyber security attacks, including jamming [2] and GPS spoofing [3, 4]. It will be enabled by cognitive radio, making these networks open to new attacks, including primary user emulation attacks [5] and spectrum sensing data falsification [6]. Thus, it is important to explore the cybersecurity implications of 5G systems [7, 8]. Jammers create an unwanted denial of service by transmitting radio signals that flood the communication channels aiming at decreasing SNR of legitimate users thereby interrupting their communication. The attacks can be easily launched using software defined radio units such as the GNU radio and universal software radio peripherals, which are cheap and easily accessible. Jammers can target any particular frequency channel with low cost [9]. Jamming attacks can be divided into four main types: constant jammers, random jammers, deceptive jammers, and reactive jammers. Constant jammers launch an attack by transmitting a continuous high-power noise sweeping from a channel to another following a fixed strategy and repeating this process over time. Random jammers operate randomly and do not follow any specific strategy jumping from a channel to another. Deceptive jammers send illegitimate packets through the wireless channels to keep them busy. Reactive jammers continuously monitor the state of the frequency channels and target only the channels used for communication [10]. In addition, jammers can be classified as: regular or smart. Regular jammers cannot sense the ongoing transmitted signals and they all play simultaneously. Smart jammers can learn quickly, sense and determine how the legitimate users are transmitting their signals and they can update their attacks' strategies or adjust the transmission power to more damage the legitimate transmission.

A number of jamming detection techniques have been proposed [11-19]. These techniques can be categorized into two main classes: non-machine learning [11-17] and machine learning based [18,19]. Non-machine learning methods perform using some parameters and strategies including threshold, fuzzy logic, game theory, channel surfing, mapping jammed region, and timing channel. In [12], the authors developed a time series model in which they measured the state of the link over series of time and compared it with the past link data to detect the state of the communication link. In [13], the authors developed a threshold based model using the packet loss, throughput, and message invalidation ratio to evaluate the performance of the wireless channels in time-critical applications. In [14], the authors proposed two timing channel based models for jamming detection. One model computes the poor packet delivery ratio based on the received signal strength while the second model computes the throughput. Based on these two parameters, they were able to detect whether the link is attacked or not. In [15], the author proposed a fuzzy logic centralized jamming detection technique based on the received signal strength, packet delivery ratio, bad packet ratio, and channel clear assessment parameters. This model also developed a base station to run the detection algorithm which computes the packet delivery to packet received ratio and the signal to noise ratio from the received data to determine the duration of this attack [16, 17].

Machine learning methods are based on classifiers like neural networks and support vector machine with different features to detect jamming attacks. For instance, the authors of [18] proposed an artificial neural network based algorithm for cyclic spectral analysis and wideband spectrum sensing. Based on the signal quality and the modulation, the algorithm distinguishes the jamming signals from the narrowband signals. In [19], the authors designed a machine learning based jamming detection system via support vector machine, adaptive boosting, and expectation maximization algorithms. Noise, busy channel ratio, packet delivery ratio, and maximum inactive time were used to detect jamming attacks. Most of the previously mentioned techniques [11-19] require more resources and ultimately serve only as a stopgap. They can detect the state of the link as down, but often they cannot identify the source of the outage of the service. In addition, these techniques have relatively high probability of false alarm. They need accurate algorithms for training and testing the classification models. Features selection and learning curves are often neglected while they are one of the most important processes in designing detection techniques with machine learning. Thus, there is a great need for efficient and fast detection techniques able to detect jamming attacks more accurately.

In this paper, we propose using machine learning to detect the transmission link state between a transmitter and a receiver to verify if it is attacked. Machine learning based models can achieve high detection accuracy if the following steps are carefully considered: selecting appropriate input features, measuring, collecting, building a large dataset, and using accurate methodology to train, validate, and test the model. Features and parameters used to detect jamming attacks are: bad packet ratio, packet delivery ratio, received signal strength, and clear channel assessment. We investigated techniques of selecting appropriate features and assessing the communication link status. We built a large dataset to train, validate, and test machine learning models. Randomization and normalization of the dataset and cross-validation techniques were performed to avoid the problem of underfitting. The rest of the paper is organized as follows. Section II describes the jamming attack model and its classification features. Section III discusses the simulation results. Finally, a conclusion is given at the end.

## II. METHODOLOGY

The jamming attacks detection in this paper is formulated as a classification problem in which the classifier has to choose between two states: the link is lost because of a jammer or the link is lost because of another reason. The reason behind using machine learning theory to solve this problem is the success of this theory to deal with complex problems within an acceptable time and using reasonable resources. Designing a successful machine learning algorithm requires the selection of appropriate features. In this work, several features were selected to identify the presence of jamming attacks. Using one parameter only is not enough to detect if there is a jamming attack. In addition, it can be complicated to find analytic relations between these parameters and the status of the link. For these reasons, machine learning theory is used to find an empiric relation among these four metrics in order to detect jamming attacks.

In this section, we describe the jamming attack model used, the feature selection and the four parameters used to detect jamming attacks. Next, we describe the machine learning techniques used, which are random forest, support vector machine with different kernels, and neural networks.

### A. Jamming attacks model

Jamming attacks target both physical layer and cross-layer of the wireless networks [20]. Under these attacks, the desired communication between a transmitter at location A and a receiver at location B is interrupted by the jamming signals which keep the channel busy. When the jamming signals occupy the channel for a longer period of time, they can create a denial of service [21]. If the desired signal, at location A, is denoted by $x(t)$ and the received signal, at location B, is denoted by $y(t)$, then this received signal at the location B, in the absence of the jammer signal, is given by

$$y(t) = x(t) + n(t) \qquad (1)$$

where $x(t)$ is the desired signal, $y(t)$ is the received signal, and $n(t)$ is an additive white Gaussian noise present within the path between the transmitter and the receiver. The jammers can counterfeit the desired signals creating a signal denoted $x_j(t)$ to flood the channel. The receive signal at location B, in the presence of the jammer signal, is then given as:

$$y(t) = x(t) + x_j(t) + n(t) + n_1(t) \qquad (2)$$

where $x_j(t)$ is the jammer signal and $n_1(t)$ is the noise within the path between the receiver and the jammer's location. Thus, the jamming detection problem can be stated as a hypothesis, in which the receiver has to choose between two states, $H_0$ and $H_a$. $H_0$ is the state where the received signal is not jammed, while $H_a$ is the state where the received signal is jammed. This problem thus can be expressed as a classification problem, in which the machine learning classifier has to attribute the incoming signal into one of the two classes: the received signal is the desired signal, class A, or the received signal is a jamming signal, class B.

### B. Feature selection

Parameters used to detect jamming attacks are bad packet ratio, packet delivery ratio, received signal strength, and clear channel assessment. The reason behind using these four parameters is that all communication systems are equipped with network interface cards that possess diagnostic mechanisms which allow the estimation of these metrics [22-26].

Bad packet ratio is one of the most important parameters to detect jamming attacks. It refers to the percentage of incorrect packages received [23]. It can be measured at the receiver end and is expressed as

$$PR = \frac{Number\ of\ erroneous\ received\ packages}{Total\ number\ of\ received\ packages} \qquad (3)$$

The receivers compute this bad packet ratio by verifying the frame check sequence of the incoming packets at the medium access control level. If the channel is under any kind of attack, the bad packet ratio increases while it is very low when the link status is good for transmission. The packet delivery ratio refers to the percentage of correctly delivered packages. It is measured at the transmitter end and expressed as

$$PDR = \frac{Number\ of\ package\ delevirey\ correctly}{Total\ number\ of\ transmitted\ packages} \quad (4)$$

The receiver sends back an acknowledgment packet to the transmitter each time it receives a correct packet. The packet delivery ratio is very high when the link status is good while its value decreases exponentially if the link is under any attack. The clear channel assessment can be used to measure the number of transmitter's attempts to send a package and the channel is found to be occupied. The value of this parameter increases if the channel is under jamming attacks.

The received signal strength, RSS, measures the surrounding power of the receiver. It is high when there is no attack; however, it decreases if the channel is under any kind of attack. RSS at the receiver can be expressed as

$$RSS = \frac{P_t * G_t * G_r * (ht^2 * hr^2)}{d^4} \quad (5)$$

where $P_t$ is the transmitter signal power, $G_t$ and $G_r$ is the gain of the antenna at transmitter and receiver respectively, $ht$ and $hr$ are the height of antenna at transmitter and receiver, and $d$ is the distance between transmitter and receiver. Equation (20) is expressed as

$$RSS = K\frac{P_t}{d^4} \quad (6)$$

where $k$ is a constant such that $k = G_t * G_r * (ht^2 * hr^2)$.

C. *Machine Learning Algorithms*

A description of each of the machine learning algorithms is given below.

*1) Random forest*

Random forest is a hierarchical classifier method composed of a large number of decision trees. In this approach, the test data is classified by sorting trees based on their feature values. Each decision tree consists of one node and several branches. The decision node is the feature of the test data to be classified, and the branches represent a value that the node can predict. The reason behind using a large number of trees is to avoid the problem of overfitting. System variance is reduced, which eventually increases the performance of the final model. Basic parameters to random forest classifier can be the total number of trees to be generated and decision tree related parameters like minimum split, and split criteria. Random forest is a predictor that collects the information from each tree $\{r_n(x, \theta_m, D_n, m \geq 1)\}$, where $\theta_1, \theta_2 \ldots \theta_m$ are the output of each random trees. These trees are combined to form the aggregated estimation to train the forest using:

$$\bar{r}_n(X, D_n) = E_\theta[r_n(X, \theta, D_n)] \quad (7)$$

where $E_\theta$ is the expectation with respect to the random parameter, conditionally, on $X$ and the data set $D_n$. Once the forest is trained, each tree can predict independently to output values using the following equation:

$$f_n^j(x) = \frac{1}{N^e(A_n(x))}\sum_{\substack{Y_i \in A_n(x) \\ I_i = e}} Y_i \quad (8)$$

where $x$ is the query point of each tree. The forest averages the predictions of each tree to get the final value:

$$f_n^{(M)}(x) = \frac{1}{M}\sum_{j=1}^{M} f_n^j(x) \quad (9)$$

where $A_n(x)$ is the leaf containing $x$ and $N^e(A_n(x))$ is the number of estimation points it contains. For the binary classification using random forest, random response $Y$ takes only two values in $\{0, 1\}$. Given $X$, random forest has to attribute 0 or 1 to $Y$. Random forest is based on Borel classification measurable rule $m_n$, which is used to estimate the label of $Y$ from $x$ and $D_n$ where the classifier $m_n$ is consistent if its conditional probability of error is low which can be expressed as follow

$$L(m_n) = P[m_n(X) \neq Y[D_n]] \quad (10)$$

$$L(m_n) = P[m_n(X) \neq Y[D_n]] \quad (11)$$

where $L^*$ is the error of the optimal but unknown and $E$ is the expectations with respect to the random parameter $\theta$. After that Bayes classifier is used to get the output for both 0 and 1.

$$m^*(x) = \begin{cases} 1, & if\ P[Y = 1, X = x] > [Y = 0, X = x < 0] \\ 0, & otherwise. \end{cases} \quad (12)$$

In the classification situation where the dataset is divided into several classes based on the input parameters threshold values, the random forest classifier is obtained via a majority vote among the classification trees, that is

$$m_{M,n}(x; \theta_1 \ldots \theta_m, D_n) = \begin{cases} 1, & if\ \frac{1}{M}\sum_{j=1}^{M} m_n(x; \theta_j, D_n) > 1/2] \\ 0, & otherwise. \end{cases} \quad (13)$$

where $n$ is the number of trees and M tends to infinity number of trees. Based on the majority votes, the output is classified as 1 or 0. If more than 50% of the total trees vote for 1 then the final prediction of random forest is 1 and if more than 50% of the total trees vote for 0 then the final prediction of random forest is considered as 0.

*2) Support vector machine*

Support vector machine creates a hyperplane to separate data into two classes. The choice of the kernel determines the separation boundary between the two classes. Different kernels can be used with this model such as linear kernel, radial basis function, quadratic, and cubic kernels. The linear kernel is defined as:

$$K(x) = w^T x + b \quad (14)$$

Linear support vector machine is formulated as solving an optimization problem as:

$$\min_{w \in R^d} \|w\|^2 + C \sum_i^N \max(0, 1 - y_i K(x_i)) \quad (15)$$

Quadratic and cubic kernels are polynomial kernels with degrees of 2 and 3, respectively. Polynomials kernels are defined as:

$$K(x, y) = (x \cdot y + 1)^d \quad (16)$$

where $x$ and $y$ are vectors of features and d is the degree of the polynomial. Radial basis function kernel is defined as:

$$K(x, y) = \exp(-\gamma \|x - y\|^2) \quad (17)$$

*3) Neural Network*

A neural network is a biological-inspired programming paradigm, which enables a machine to learn from observational data. This network has shown a great ability to learn and solve various problems in different research areas such as image processing, signal processing, and wireless communication. A neural network consists of one input layer, one or several hidden layers, and an output layer. Each layer consists of either one or several neurons. A neuron consists of an activation function and several links connecting them to other neurons in different layers. An initial weight is associated with each link and the neural network in the learning phase try to find the set of optimal weight that minimizes the error between the hypothesis function and the given dataset labels. Each neural network consists of two main concepts, which are the forward propagation and backpropagation. Forward propagation is the simplest type of artificial neural networks where the information moves in only one direction, from input to the output through hidden layers. Input features can be denoted as $x_1, x_2, \ldots x_n$ and the input layer can be summarized by the following equations:

$$a_{(j)}^{(i)} = x_i \quad (18)$$

where $a_{(j)}^{(i)}$ is the input layer and $x_i$ is the input features. Input layer is connected with the hidden layers which can be denoted as follows: In the i[th] hidden layer, we have

$$z^{(i)} = \theta^{(i)} a^{(i)} \quad (19)$$

where $z^{(i)}$ is the hidden neuron, $\theta^{(i)}$ is each layer matrix weight, and $a^{(i)}$ is the hidden layer for the i[th] hidden layer. The final layer is the output layer which can be denoted as:

$$a^{(i)} = g(z^{(i)}) \quad (20)$$

Neural network cost function is used to find out the optimal output based on different number of layers and neurons. It is expressed by:

$$J(\theta) = -\frac{1}{m} \sum_{i=1}^{m} \sum_{k=1}^{k} [y_k^{(i)} \log\left(\left(h_\theta(x^{(i)})\right)_k\right) + (1 - y_k^{(i)}) \log(1 - \left(h_\theta(x^{(i)})\right)_k)] + \frac{\lambda}{2m} \sum_{l=1}^{L-1} \sum_{i=1}^{s_l} \sum_{j=1}^{s_{l+1}} (\theta_{j,i}^{(l)})^2 \quad (21)$$

where $J(\theta)$ is the cost function, $h_\theta$ is the hypothesis function, and $\lambda$ is the regularization factor. Regularization cost function is used to reduce the effect of over bias and under bias by regularization factor ($\lambda$).

$$J(\theta) = \frac{1}{m} \sum_{i=1}^{m} \sum_{k=1}^{k} [-y_k^{(i)} \log\left(\left(h_\theta(x^{(i)})\right)_k\right) - (1 - y_k^{(i)}) \log(1 - \left(h_\theta(x^{(i)})\right)_k)] + \frac{\lambda}{2m} [\sum_{l=1}^{L-1} \sum_{i=1}^{s_l} (\theta_{j,k}^{(1)})^2 + \sum_{j=1}^{L} \sum_{k=1}^{s_l} (\theta_{j,k}^{(2)})^2] \quad (22)$$

where the regularization factor $\lambda$ is used to reduce the effect of the over bias and under bias. Neural networks use different optimization techniques such as Adam, gradient descent, and stochastic gradient descent to minimize the cost function. At the output layer, neural network uses a sigmoid function to attribute the new dataset into one of the two classes, in the case of binary classification, by calculating the hypothesis using the weights determined in the leaning parts. If this hypothesis is greater than 0.5, than it concludes that "y=1"; otherwise "y=0".

III. RESULTS AND DISCUSSION

To validate the machine learning models, four different parameters were used as features to detect jamming attacks. A real environment simulation was performed to collect measurements of these parameters in the two scenarios: link is under attack and link under no attack. To train and test the models, the dataset was divided into *N* folds using the cross-validation technique. *N*-fold cross-validation was used to divide the dataset into *N* number of subsamples with equal size. In this work, machines learning algorithms were trained with a different number of fold sizes such as 2, 5, 10, and 20 to evaluate the performance of these machines for the given dataset. For instance, if the number of folds is 10, then the total data is divided into 10 folds and randomly the algorithm selects 9 folds to train the model and 1 fold is used to test the machine. This process is repeated until the dataset is tested on the 10 folds. To evaluate the performance of the classifiers, several metrics were used, namely probabilities of detection, false alarm, miss detection, and accuracy. $P_d$ refers to the likelihood that the detection technique attributes signals coming from a jammer to the class of jamming signals meaning that it correctly detects that the link is under a jamming attack

$$P_d = \frac{\text{Number of truly detected attacks}}{\text{Total number of attacks}} \quad (23)$$

$P_{md}$ is the percentage of attacks that the algorithm miss detected. It is given by:

$$P_m = \frac{\text{Number of miss detected attacks}}{\text{Total number of attacks}} \quad (24)$$

$P_{fa}$ is the percentage of non-attacks that the algorithm detected as attacks

$$P_{fa} = \frac{\text{Number of non-attacks detected as an attack}}{\text{Total number of non-attacks}} \quad (25)$$

Accuracy gives the total number of attacks and non-attacks that are detected accurately compared to the total number of trials. It is given by:

$$Accuracy = \frac{\text{Total number of correctly detected attack and non-attack trials}}{\text{Total number of trials}} \quad (26)$$

We conducted several experiments and examples of results are given from Fig. 1 to Fig. 4. Fig. 1 shows the accuracy of the classification using random forest versus the number of estimators for a different number of folds in the cross-validation. It can be seen from this figure that for all values of K-folds, 5, 10, and 20, the accuracy of the classification is exponentially increasing as a function of the number of estimators, for numbers of estimators less than 60. However, this accuracy remains slightly constant for numbers higher than 60 and in some cases, it drops. This figure also shows the impact of the number of folds on the accuracy. One can see that with 20 folds, the accuracy is higher than the one with 10 and 5 folds.

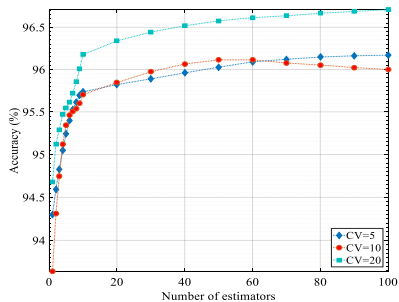

Fig. 1. Accuracy versus the number of estimators for random forest using different number of k-folds cross validation CV=5, 10, and 20.

To select the best support vector model, we investigated how the accuracy varies for different kernels and regularization parameter $C$ in order to select the best combination for this given dataset. Fig. 2 shows the accuracy of the classification function of the regularization factor "$C$" for linear, quadratic, cubic, radial basis function, and sigmoid kernels. From this figure, it can be seen that the impact of the regularization factor "$C$" does not change the accuracy very much. However, as one can see the choice of the kernel impacts the accuracy. The accuracy is high for radial basis function kernel, followed by linear, cubic, sigmoid, and then quadratic kernels. Support vector machine with radial basis function and regularization factor equal to 3 has the highest accuracy of 94%.

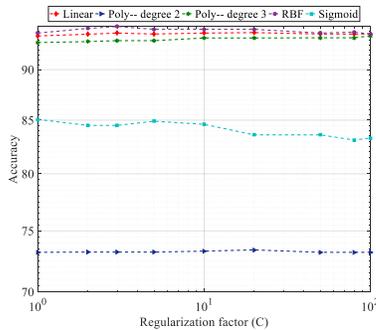

Fig. 2. Accuracy Vs regularization factor for support vector machine.

Fig. 3 shows the accuracy of the classification function of the number of hidden neurons in one hidden layer of neural network for a different number of k-folds cross-validation. One can observe that the impact of the number of hidden neurons and the number of cross-validation technique is not significant as the accuracy of the classification remains around 94%, but the highest one is achieved with 1 neuron with 5-folds cross-validation and with 100 neurons with 10-folds cross-validation.

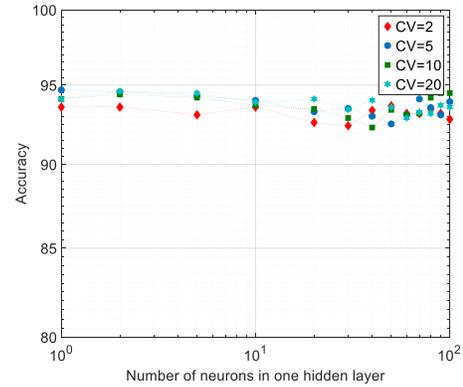

Fig. 3. Accuracy of neural network Vs number of neurons in one hidden layer.

Fig. 4 shows $P_d$ Vs $P_{fa}$ using linear, polynomial with degree 2, radial basis function SVM, neural network with two hidden layers of 2 neurons each, and random forest with 100 estimators. One can see that $P_d$ increases as $P_{fa}$ increases. In addition, it can be observed that random forest has the higher ROC followed by radial basis function SVM, cubic SVM, linear SVM, and then the neural network which means that random forest outperforms other algorithms regarding the ROC curve.

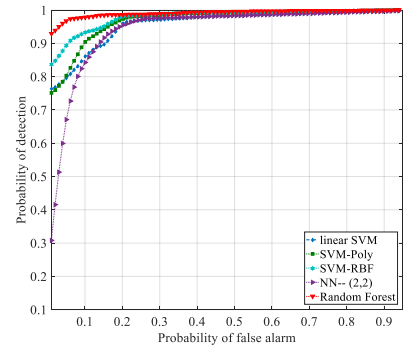

Fig. 4. Probability of detection Vs the probability of false alarm.

Table I compares the performance of the jamming detection techniques based on machine learning classifiers for the four evaluation metrics. Random forest achieves the highest probability of detection with 97.5 %, followed by cubic SVM with 97.1%, neural network with 96.4%, linear SVM with 86.9%, RBF SVM with 86.2%, sigmoid SVM with 73.8%, and quadratic SVM with 72 %. It can also be seen that random forest has the lowest probabilities of false alarm of 5.6%, followed by neural network with 11.1%, RBF SVM with 27.2%, linear SVM with 27.25%, sigmoid SVM with 40.1%, cubic SVM with 54%, and quadratic SVM with 65.2%. This table shows also that random forest has the lowest miss detection with 2.5%, followed by cubic SVM with 2.9%, neural network with 3.6%, linear SVM with 13.1%, RBF SVM with 13.8%, sigmoid SVM with 26.22%, and quadratic SVM with 28%. In terms of accuracy, random forest has an accuracy as high as 96.6% followed by neural network with 94.4%, RBF SVM with 84.7%, linear SVM with 83%, sigmoid SVM with 82.6%, cubic SVM 70.1%, quadratic SVM with 62%.

TABLE I. PERFORMANCE COMPARISON

| Classification technique | $P_d$ (%) | $P_{fa}$ (%) | $P_{md}$ (%) | Accuracy (%) |
|---|---|---|---|---|
| Linear SVM | 86.9 | 27.25 | 13.1 | 83 |
| Quadratic SVM | 72 | 65.2 | 28 | 62 |
| Cubic SVM | 97.1 | 54 | 2.9 | 70.1 |
| RBF SVM | 86.2 | 27.2 | 13.8 | 84.7 |
| Sigmoid SVM | 73.8 | 40.1 | 26.22 | 82.6 |
| Neural network | 96.4 | 11.1 | 3.6 | 94.4 |
| Random Forest estimators = 100 | 97.5 | 5.6 | 2.5 | 96.6 |

CONCLUSION

5G technology is designed to be resilient to jamming attacks by using millimeter wave band. However, it is also designed to use frequencies below 6 GHz, which are easy to target by jammers. Smart jamming detection techniques are required to prevent these attacks. In this paper, we reviewed the existing jamming detection techniques. We investigated and compared the performance of several machine learning models to detect jamming attacks. Feature extraction and feature selection were performed and a large dataset was constructed to train, validate, and teste random forest, support vector machine, and neural network algorithms. We used a cross-validation technique and provided learning curves to evaluate the performance of these models based on a number of metrics. The results show that random forest based technique detects jamming attacks with a very high accuracy and a low cost. $P_d$ of random forest based detection is as high as 97.5% whereas $P_{fa}$ of the neural network and cubic support vector machine is around 96.4% and 97.1%. $P_{md}$ and $P_{fa}$ of random forest are also very low compared to neural network and cubic support vector machine which are 5.6% and 2.5%. High $P_d$ and low $P_{fa}$ make this proposed model suitable for jamming attack detection. These trained machines are able to process a huge number of data within a very short time, which helps increase efficiency and reduce the processing time. Future work includes investigating the efficiency of deep learning in detecting all types of jamming attacks.